\documentclass[pra,aps,twocolumn,nobalancelastpage,superscriptaddress]{revtex4-1}

\usepackage[utf8]{inputenc}
\usepackage[T1]{fontenc}
\usepackage[english]{babel}
\usepackage{csquotes}
\usepackage{comment}

\usepackage{amsmath}
\usepackage{relsize}

\usepackage[margin=2cm]{geometry}

\usepackage{siunitx}
\DeclareSIUnit{\sample}{Sa}
\DeclareSIUnit{\baud}{Baud}
\DeclareSIUnit{\snu}{SNU}

\usepackage{graphicx}
\graphicspath{{./figures/}}
\usepackage{pgf}
\usepackage{float}
\usepackage{array}
\usepackage{multirow}

\usepackage[hidelinks]{hyperref}
\usepackage{natbib}
\usepackage{glossaries-extra}
\setabbreviationstyle[acronym]{long-short}
\newacronym{cvqkd}{CV-QKD}{Continuous Variable Quantum Key Distribution}
\newacronym{cnrs}{CNRS}{Centre National de la Recherche Scientifique}
\newacronym{c2n}{C2N}{Centre de Nanosciences et de Nanotechnologies}
\newacronym{cea}{CEA}{Commissariat à l'Énergie Atomique et aux énergies alternatives}
\newacronym{leti}{LETI}{Laboratoire d'Électronique et de Technologie de l'Information}
\newacronym{voa}{VOA}{Variable Optical Attenuators}
\newacronym{tia}{TIA}{Transimpedance Amplifier}
\newacronym{lo}{LO}{Local Oscillator}
\newacronym{spice}{SPICE}{Simulation Program with Integrated Circuit Emphasis}
\newacronym{pcb}{PCB}{Printed Circuit Board}
\newacronym{skr}{SKR}{Secret Key Rate}
\newacronym{dsp}{DSP}{Digital Signal Processing}
\newacronym{rbw}{RBW}{Resolution Bandwidth}
\newacronym{vbw}{VBW}{Video Bandwidth}
\newacronym{psd}{PSD}{Power Spectral Density}
\newacronym{psds}{PSDs}{Power Spectral Densities}
\newacronym{ossb}{OSSB}{Optical Single Sideband}
\newacronym{psk}{PSK}{Phase-Shift Keying}
\newacronym{qam}{QAM}{Quadrature-Amplitude Modulation}
\newacronym{pcs}{PCS}{Probabilistic Constellation Shaping}
\newacronym{rrc}{RRC}{Root Raised Cosine}
\newacronym{rc}{RC}{Raised Cosine}
\newacronym{cazac}{CAZAC}{Constant Amplitude Zero Autocorrelation Waveform}
\newacronym{pic}{PIC}{Photonic Integrated Circuit}
\newacronym{mbc}{MBC}{Modulator Bias Controller}
\newacronym{adc}{ADC}{Analog-to-Digital Converter}
\newacronym{dac}{DAC}{Digital-to-Analog Converter}
\newacronym{isi}{ISI}{Intersymbol Interference}
\newacronym{rnl}{RNL}{Relative Non-Linearity}
\newacronym{dnl}{DNL}{Dimensionnal Non-Linearity}
\newacronym{qkd}{QKD}{Quantum Key Distribution}
\newacronym{dvqkd}{DV-QKD}{Discrete Variable Quantum Key Distribution}
\newacronym{otp}{OTP}{One-Time Pad}
\newacronym{snspd}{SNSPD}{Superconducting Nanowire Single-Photon Detectors}
\newacronym{qrng}{QRNG}{Quantum Random Number Generator}
\newacronym{fpga}{FPGA}{Field Programmable Gate Array}
\newacronym{ldpc}{LDPC}{Low-Density Parity-Check Code}
\newacronym{snu}{SNU}{Shot Noise Units}
\newacronym{bhd}{BHD}{Balanced Homodyne Detectors}
\newacronym{snr}{SNR}{Signal-to-Noise Ratio}
\newacronym{fer}{FER}{Frame Error Rate}

\usepackage{circuitikz}

\begin{document}

\title{Experimental demonstration of Continuous-Variable Quantum Key Distribution with a silicon photonics integrated receiver}

\author{Yoann Piétri}
\affiliation{Sorbonne Université, CNRS, LIP6, 75005 Paris, France}

\author{Luis Trigo Vidarte}
\affiliation{ICFO-Institut de Ciencies Fotoniques, The Barcelona Institute of Science and Technology, Castelldefels (Barcelona) 08860, Spain}

\author{Matteo Schiavon}
\affiliation{Sorbonne Université, CNRS, LIP6, 75005 Paris, France}

\author{Laurent Vivien}
\affiliation{Université Paris-Saclay, CNRS, Centre de Nanosciences et de Nanotechnologies (C2N), 91120 Palaiseau, France}

\author{Philippe Grangier}
\affiliation{Université Paris-Saclay, Institut d’Optique Graduate School, CNRS, Laboratoire Charles Fabry, 91127 Palaiseau, France}

\author{Amine Rhouni}
\affiliation{Sorbonne Université, CNRS, LIP6, 75005 Paris, France}

\author{Eleni Diamanti}
\affiliation{Sorbonne Université, CNRS, LIP6, 75005 Paris, France}

\begin{abstract}
      \glsxtrfull{qkd} is a prominent application in the field of quantum cryptography providing information-theoretic security for secret key exchange.  The implementation of QKD systems on photonic integrated circuits (PICs) can reduce the size and cost of such systems and facilitate their deployment in practical infrastructures. To this end, continuous-variable (CV) QKD systems are particularly well-suited as they do not require single-photon detectors, whose integration is presently challenging.
      Here we present a CV-QKD receiver based on a silicon PIC capable of performing balanced detection. We characterize its performance in a laboratory QKD setup using a frequency multiplexed pilot scheme with specifically designed data processing allowing for high modulation and secret key rates. The obtained excess noise values are compatible with asymptotic secret key rates of $2.4\,\si{\mega\bit\per\second}$ and $220\,\si{\kilo\bit\per\second}$ at an emulated distance of $10\,\si{\kilo\meter}$ and $23\,\si{\kilo\meter}$, respectively. These results demonstrate the potential of this technology towards fully integrated devices suitable for high-speed, metropolitan-distance secure communication.
\end{abstract}

\date{\today}
\maketitle

\section{Introduction}

\gls{qkd} is a cryptographic task enabling two trusted users, usually referred to as Alice and Bob, who have access to an unsecure quantum channel and a public but authenticated classical channel, to exchange a random string of bits whose security is ensured by the laws of physics. This means that a potential eavesdropper has a negligible amount of information on the final string of bits, the secret key~\cite{Pirandola_2020}. This task combined with the use of encryption schemes such as the \gls{otp}~\cite{Shannon_1949} allows for information-theoretic, long-term security for secret message exchange.

The two main families of QKD protocols rely on encoding the information on discrete degrees of freedom of single photons, such a polarization~\cite{Bennett_1984}, in so-called discrete-variable (DV) QKD, or on continuous degrees of freedom of light, such as the quadrature components of coherent states~\cite{Grosshans_2002}, in CV-QKD. Driven by the need to protect the transmission of sensitive data against attacks enabled by rapid technological advancements, significant progress has been achieved in QKD systems from both families in the last years~\cite{Pirandola_2020,Xu_rmp2020}. This includes proposals of important variants, such as measurement device independent and twin-field QKD~\cite{Lo_2012,Lucamarini_2018}, records in secret key generation distance and rate~\cite{Boaron_prl2018,Zhang_2020,Roumestan_arXiv2022,Wang_2022}, and demonstrations or feasibility studies over satellite links~\cite{Liao2017,Dequal_2021}. 

\begin{figure*}[!t]
\centering
    \includegraphics[scale=0.18]{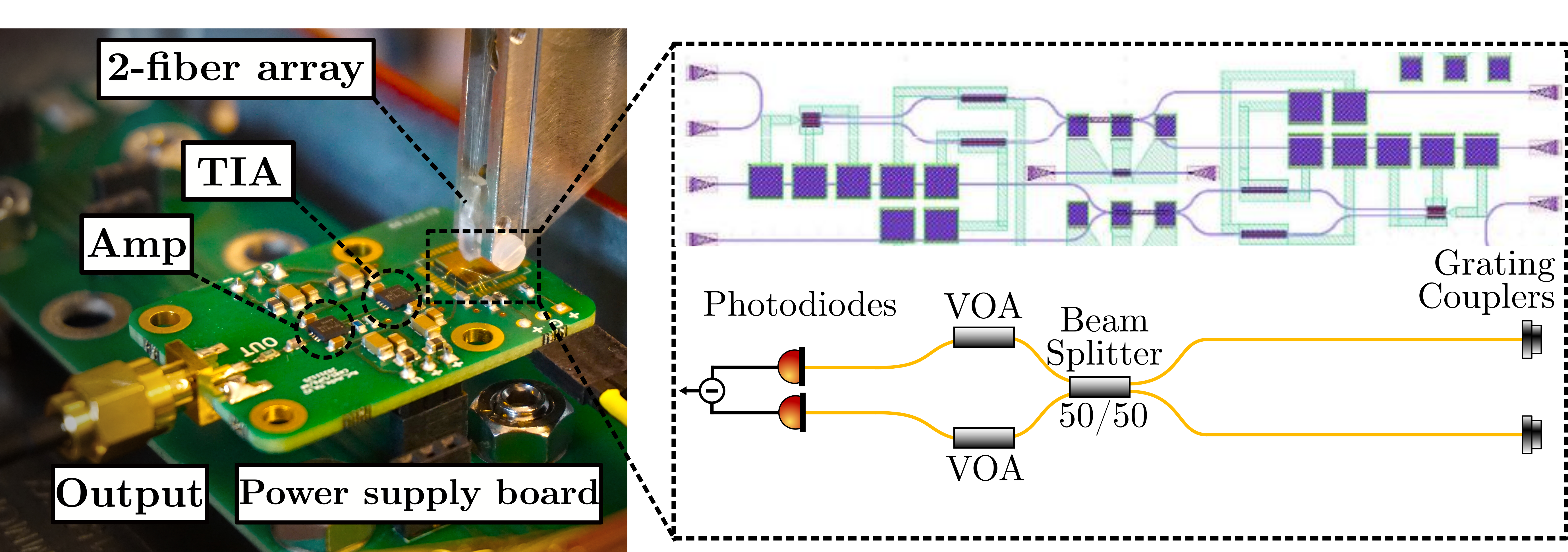}
    \caption{\textbf{The CV-QKD Si PIC receiver platform.} On the left, a picture of the platform. The photonic integrated chip (PIC) is wirebonded to the printed circuit board (PCB), which is itself on its power supply board. Optical coupling is done through a fiber array with two fibers on a 5-axis mechanical stage. The transimpedance and voltage amplifiers and the output of the device are also shown. On the right, the layout of the area of interest in the PIC and the associated schematic view, showing two grating couplers, one 50/50 beam splitter, two variable optical attenuators and two photodiodes. The chip hosts other photonic functions that are beyond the scope of this paper.}
    \label{fig:receiver}
\end{figure*}

Several challenges towards improving the performance and practical deployment of QKD systems in terms of key rate, distance, practical security, cost, size, stability or satellite communication may be effectively addressed with photonic integration. For this reason, significant efforts have been put in this direction, for both the transmitter and receiver devices. For DV-QKD systems, which enjoy maturity and good loss tolerance, the integration of transmitters continues to progress with very promising results~\cite{Sax_arXiv2022,Dolphin_arXiv2023}, but there is an important gap with the development of integrated receivers~\cite{Liu2022}. This is mainly due to the difficulty in integrating avalanche photodiodes and \gls{snspd} with other photonic circuits. Even with the development of technologies such as waveguide-integrated \gls{snspd}~\cite{FerrariSchuckPernice2018}, the cooling requirement is an issue since those detectors only achieve their targeted sensitivity at (sub)Kelvin temperatures. CV-QKD systems on the other hand benefit from an operation that only requires coherent detectors with standard room-temperature photodiodes, which opens the way to a high overall level of integration. Although this still remains challenging, some promising demonstrations of integrated circuits capable of performing homodyne detection~\cite{Bruynsteen_2021, Raffaelli_2018} and applications to CV-QKD~\cite{Zhang_2019, Hajomer_2023} have already been shown. In~\cite{Zhang_2019}, silicon (Si) photonic integrated devices are used for both the transmitter and the receiver. The implemented scheme is based on a pulsed configuration, with a transmitted phase reference (or local oscillator) and homodyne detection. % and sifting. 
More recent works tend to implement CV-QKD protocols with heterodyne detection and advanced pulse shaping to maximise the bandwidth usage, such as in the work in~\cite{Hajomer_2023} where the authors used a Si photonic integrated receiver that performs phase-diverse heterodyne detection, with two balanced detectors. While our frequency scheme is similar to the one in this work, our receiver is based on a single balanced detector in an RF-heterodyne configuration.

More specifically, in this work we present and characterize a receiver platform based on a Si PIC and we report on its use in a CV-QKD setup with Gaussian modulated states and Optical Single Sideband modulation. We apply a scheme that optimizes bandwidth utilisation similar to the work in~\cite{Jain_2022}, featuring specifically designed data processing algorithms, with shaped and frequency displaced quantum symbols, and frequency multiplexed pilots, allowing for high baud rates for the quantum symbols, and hence higher key rates. In section~\ref{sec:receiver_platform}, we describe the receiver platform and in particular the PIC (subsection~\ref{subsec:pic}) and its amplification circuit (subsection~\ref{subsec:amplification_circuit}). Measured performances of the receiver as a standalone device for CV-QKD are provided in section~\ref{sec:receiver}, while in section~\ref{sec:cvqkd}, we demonstrate the achieved performance when it is used into a full CV-QKD setup. We conclude in section~\ref{sec:concl} with further challenges and perspectives.

\section{Presentation of the receiver platform\label{sec:receiver_platform}}

The receiver platform is composed of two main parts, as shown in Fig.~\ref{fig:receiver}: the Si photonic integrated circuit and the amplification chain circuit. The PIC was fabricated at CEA/Leti and includes several functions: grating couplers, splitters, \glsxtrfull{voa} and germanium photodiode detectors. The layout is zoomed on the area of interest showing the photonic components discussed in this paper. The receiver is mounted on a power supply board capable of hosting up to four boards.   

\subsection{Photonic Integrated Circuit \label{subsec:pic}}

The PIC hosts four \gls{bhd} accessible through input grating couplers. In this application, we use a single \gls{bhd}.   

The signal carrying the quantum information and the local oscillator (LO) are injected using a fiber array through a pair of grating couplers, with a pitch of $127\,\si{\micro\meter}$. Both travel through waveguides until they are mixed in a 50/50 beam splitter acting as a 180$^\circ$ hybrid mixer. Two germanium photodiodes connected in series receive the mixed outputs through the VOAs. These allow to balance the optical power in both arms, which is critical to avoid saturation and non-linearity. They are driven by applying an external voltage making use of the free carrier plasma dispersion effects to change the waveguide absorption and reflection coefficients~\cite{vivien_2016_handbook}. 

\begin{figure}[htbp]
    \centering
    \input{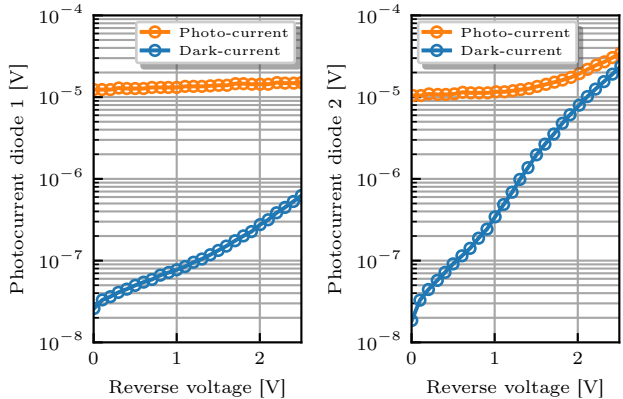}
    \caption{\textbf{I-V characteristics of the balanced photodetector}. The curves correspond to the two photodiodes shown in Fig.~\ref{fig:receiver} under dark and illumination conditions. To keep a low dark current and hence a low electronic noise, we choose a reverse bias of $0.5\,\si{\volt}$.}
    \label{fig:c2n_characterization}
\end{figure}

We performed on-chip static current voltage (I-V) measurements using an electrical probe station for voltage biasing and a single mode fiber for optical coupling under dark and illumination conditions (at a wavelength $\lambda = 1550\,\si{\nano\meter}$). Fig.~\ref{fig:c2n_characterization} shows the results for each photodiode of the \gls{bhd}. The dark current increases as a function of the reverse voltage while the photocurrent remains almost constant up to $1.5\,\si{\volt}$. The difference in the dark current contribution for each photodiode is due to dislocation. To guarantee a high photo-to-dark current ratio and for a better common mode rejection, we set the reverse voltage at $0.5\,\si{\volt}$ for all measurements in this work.

\subsection{Amplification chain circuit \label{subsec:amplification_circuit}}

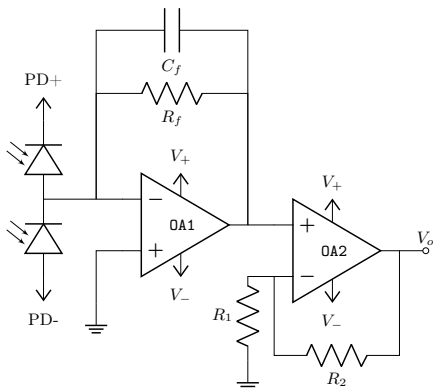
\begin{figure}
    \centering
    \scalebox{0.7}{\begin{circuitikz}[]
    \draw (0,0) node[vcc](VCC){PD+} to[pD,mirror,invert] ++(0,-1.5) node[](junc){} to [pD,mirror,invert] ++ (0,-1.5) node[vee]{PD-};
    \draw (junc-|VCC) to ++ (1,0) node[](restia){} to ++ (0.5,0) node[op amp, noinv input down, anchor=-](OA1){\texttt{OA1}};
    \draw (OA1.up) -- ++(0, 0.01) node[vcc]{$V_+$};
    \draw (OA1.down) -- ++(0, -0.01) node[vee]{$V_-$};
    \draw (OA1.+) to ++(-0.5,0) to ++(0,-1) node[ground]{}; 
    \draw (OA1.out) to ++(0,2.5) node[](rfstart){} to[R=$R_f$] (rfstart -| restia) to (junc-|restia);
    \draw (OA1.out) to ++(0,3.7) node[](cfstart){} to[C=$C_f$] (cfstart -| restia) to (junc-|restia);
    \draw (OA1.out) to ++ (0.5,0) node[op amp, noinv input up, anchor=+](OA2){\texttt{OA2}};
    \draw (OA2.up) -- ++(0, 0.01) node[vcc]{$V_+$};
    \draw (OA2.down) -- ++(0, -0.01) node[vee]{$V_-$};
    \draw (OA2.out) to [short, -o] ++(0.5,0) node[above]{$V_o$};
    \draw (OA2.-) to ++(-0.5,0) to[R, l_=$R_1$] ++(0,-1.5) node(r1){} to ++(0,-0.1) node[ground](g){};
    \draw (OA2.out) to (r1-|OA2.out) to[R=$R_2$] (r1-|OA2.-) to ++(0, 1.5);
    \end{circuitikz}}
    \caption{\textbf{Principle scheme of the amplification chain.} The circuit is composed of a transimpedance amplifier followed by a non-inverting voltage amplifier. PD+ and PD- are the reverse bias voltages of the photodiodes. The power supply voltage filters and the 50 Ohms output DC filter are not shown.}
    \label{fig:amplification_circuit}
\end{figure}

The main function of the \gls{bhd} is to provide a current proportional to the difference between the two incident light powers, leading to common mode suppression. Indeed, since the signal and the local oscillator are mixed on the 50/50 beamsplitter and the \gls{bhd} is balanced thanks to the VOAs, the resulting output current is very weak. Thus, analog signal conditioning (amplification and filtering) before sampling is necessary through first a \gls{tia} and then a voltage amplifier. The methodology for the design of the TIA for our PIC was driven by the optimization of the principal parameters: signal-to-noise ratio (SNR), bandwidth and stability (all related and sensitive to parasitic components). We also emphasize the importance of the output impedance matching and the supply voltage filter in obtaining the best noise performance.

We studied several \gls{tia} architectures based on RF amplifiers; see appendix~\ref{appendix:TIA} for more details. The tests of the electronics were performed both alone, using electronic test inputs, and with a bulk \gls{bhd} made of photodiodes (Hamamatsu) before a final test with the PIC. After a series of comparative tests, we selected the OPA818 (Texas Instruments) for its best trade-off between gain-bandwidth product and SNR. We set the \gls{tia} minimum acceptable specifications for our system to be a closed-loop $Fc_{-3dB}$ bandwidth = $100\,\si{\mega\hertz}$, with a clearance (shot noise to electronic noise ratio) of at least $10\,\si{\deci\bel}$.  

The scheme of the \gls{bhd} is given in Fig.~\ref{fig:amplification_circuit}. The first stage is the TIA, whose gain is defined by $R_f$, and the minimum phase margin for stability is ensured by the compensation capacitor $C_f$. The second stage is the non-inverting voltage amplifier whose gain is set by $R_1$ and $R_2$. The I-V characteristics discussed previously have shown that PD+= $0.5\,\si{\volt}$ and PD-= $-0.5\,\si{\volt}$, while reducing the junction capacitance, guarantee the optimum bias point for an identical responsivity at low dark currents for both photodiodes of the \gls{bhd}. Following a thorough iterative procedure between simulations and fine tuning tests, we optimized the gain and capacitance values of the amplification chain circuit; see appendix~\ref{appendix:TIA} for more details.  

Fig.~\ref{fig:PICvsBulkNoise} shows the \gls{bhd} noise performance for the two designed receivers, namely the bulk detector with the Hamamatsu photodiodes ($C_f=250\,\si{\femto\farad}$) and the PIC receiver plateform ($C_f= 200 fF$). The significant improvement of the frequency bandwidth and the SNR that we observe for the PIC receiver was achieved due to the choice of a wire bonding solution of the PIC instead of a package and to a meticulous routing on PCB, which suppress almost all parasitic components. We estimate the resulting total equivalent input capacitance to be in the range of $1.8\,\si{\pico\farad} - 2\,\si{\pico\farad}$.  

\begin{figure}[hbt!]
    \centering
    \input{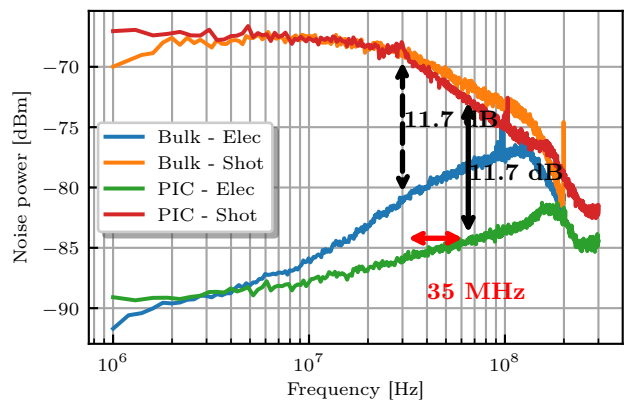}
    \caption{\textbf{Noise performance comparison between PIC and bulk BHDs}. The curves show a significant increase in the frequency bandwidth for an identical TIA between a solution with bulk photodiodes from Hamamatsu and a second with integrated photodiodes. In the second case, the parasitic capacitance at the input of the TIA is considerably reduced with respect to the test with bulk photodiodes due to wire bonding. In the working frequency range, the clearance has therefore been improved.}
    \label{fig:PICvsBulkNoise}
\end{figure}

\section{Performance of the CV-QKD receiver platform\label{sec:receiver}}

The performance of the receiver platform was first evaluated as a standalone device checking that it is linear and shot noise limited. We also measured the clearance and efficiency, which are necessary for calculating the secret key rate (SKR) of the CV-QKD exchange. Finally, we measured the frequency bandwidth of the system, which is required to bound the symbol rate and the parameters of the \gls{dsp}, which is, as we will see later, crucial for the optimisation of the system performance.

A laser at a wavelength of $1550\,\si{\nano\meter}$ was used to characterize the receiver. The injected power to the chip was manually controlled with an external VOA and a 1-to-2 50/50 beam splitter, with one output to the chip and the other to an optical power meter. A two channel voltage power was used to reverse bias and to monitor the photocurrent of each photodiode. The reverse bias voltage was set to $0.5\,\si{\volt}$. A second low noise power supply source was used to power the amplification circuit through adapted on-board filters with $+5\,\si{\volt}$ and $-5\,\si{\volt}$.  The two on-chip VOAs were driven with individual variable voltages that can be tuned between $0\,\si{\volt}$ and $+5\,\si{\volt}$. The output of the receiver wass connected to a spectrum analyser (Rohde\&Schwarz FPL1003).

We fist determined the receiver's efficiency by measuring its responsivity as follows: 
\begin{equation}
    \eta = \frac{1}{1.25}\frac{I_+ + I_-}{P_{LO}},
\end{equation}
where $P_{LO}$ is the LO power and $1.25\,\si{\ampere\per\watt}$ is the maximal responsivity at $1550\,\si{\nano\meter}$.
The VOAs were not polarized, in order to get the maximal reachable efficiency, and the responsivity was measured with a linear regression over an acquisition of several input powers, for both inputs. For the best receiver among the devices that we tested, the efficiency from the input of the fiber array to the detection was around $26\%$.

We then measured the linearity, bandwidth and the electronic-to-shot-noise ratio which are the three more important characteristics along with the efficiency for a CV-QKD receiver. For the latter parameter, the clearance was actually measured, which differs from the electronic-to-shot-noise ratio in the following sense:
\begin{equation}
    \begin{split}
        \text{electronic-to-shot-noise ratio} = V_{el} = \frac{\sigma_{el}^2}{\sigma_0^2}\\
        \text{clearance} = \frac{\sigma_{0}^2+\sigma_{el}^2}{\sigma_{el}^2} \simeq \frac{1}{V_{el}},
    \end{split}
\end{equation}
where $\sigma_{el}^2$ and $\sigma_0^2$ are the noise variance of the electronic noise and shot noise, respectively, and the approximation is true in the regime where $\sigma_{el}^2 << \sigma_{0}^2$. The clearance is typically given in $\si{\deci\bel}$ and can be found by substracting the noise power under illumination by the noise power with no illumination. However the electronic and shot noise are actually frequency-dependent; this is why we will present part of these results as Power Spectral Densities (PSDs), and these equations are then true when we integrate over the frequency spectrum. For the CV-QKD protocol, the considered spectrum is the bandwidth where the signal lies.

\begin{figure}[hbt!]
    \centering
    \input{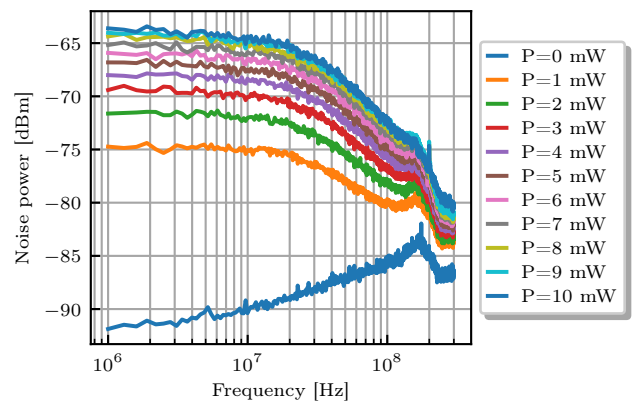}
    \caption{\textbf{Noise Power Spectral Densities (PSDs) for several input \gls{lo} power levels.} The power is given at the input of the receiver. The result for $P=0\,\si{\milli\watt}$ corresponds to the electronic noise.}
    \label{fig:psds}
\end{figure}

To estimate the three characteristics, the on-chip VOAs were tuned to reach the best common mode suppression (and prevent saturation of the receiver) and the noise power density was acquired for different input powers (including no input power for electronic noise). Those noise densities are plotted in Fig.~\ref{fig:psds}. 

As expected, the noise power increases with respect to the input optical power and should increase linearly. The linearity can be checked by plotting the integration of this noise as a function of the input power, as shown in Fig.~\ref{fig:linearity}. Here we see that the noise varies linearly until we reach an input power of about $8\,\si{\milli\watt}$. Knowing the global efficiency, we can estimate that the saturation happens at approximately $1.1\,\si{\milli\watt}$ of received power per photodiode.

\begin{figure}[hbt!]
    \centering
    \input{figures/linearity.pgf}
    \caption{\textbf{Noise integration vs. input \gls{lo} power.} The integration bandwidth is the bandwidth of the spectrum acquisition, \textit{i.e} $300\,\si{\mega\hertz}$. The linearity is ensured up to $8\,\si{\milli\watt}$ of optical input power for this detector. The linear fit in the figure corresponds to the experimental values until $8\,\si{\milli\watt}$. In this range the dimensional non-linearity was $0.02\times 10^{-9} \si{\volt\squared\per\hertz}$, and relative non-linearity was 0.80\%.}
    \label{fig:linearity}
\end{figure}

\begin{figure}
    \centering
    \input{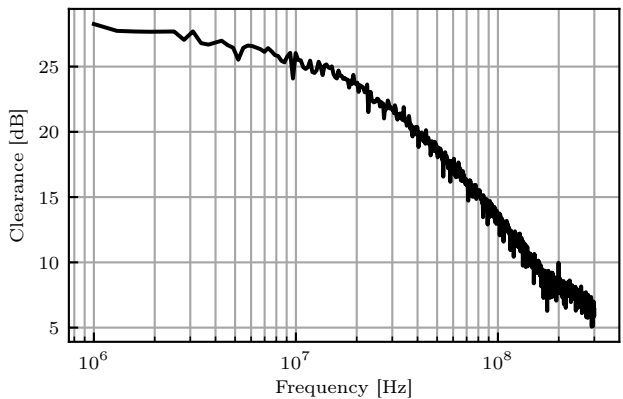}
    \caption{\textbf{Clearance vs. frequency.} The clearance decreases with the frequency as the electronic noise increases. It is around $26\,\si{\deci\bel}$ at $10\,\si{\mega\hertz}$, $14\,\si{\deci\bel}$ at $100\,\si{\mega\hertz}$, $9\,\si{\deci\bel}$ at $200\,\si{\mega\hertz}$ and $6\,\si{\deci\bel}$ at $300\,\si{\mega\hertz}$.}
    \label{fig:clearance}
\end{figure}

Knowing the linearity range, we can take the last PSD for which the linearity is ensured, and compute the clearance by calculating the ratio (or difference if in log-scale) with the electronic noise PSD. The clearance depends on the frequency as shown in Fig.~\ref{fig:clearance}. We reach a high clearance at low frequencies, around  $26\,\si{\deci\bel}$ at $1\,\si{\mega\hertz}$, and then it decreases with the frequency. The frequency bandwidth of the receiver is defined relatively to this clearance. As a clearance greater than $10\, \si{\deci\bel}$ allows in general for CV-QKD operation, we choose this threshold to define the bandwidth, which is thus around $150\,\si{\mega\hertz}$ for our PIC receiver. However, the spectrum up to $250\,\si{\mega\hertz}$ can be used also to place the classical signals for recovering the clock, the frequency and the phase (the pilot tones), since they can withstand more electronic noise (relative to the shot noise) than the quantum signal.

\section{CV-QKD system performance\label{sec:cvqkd}}

\subsection{Description of our CV-QKD scheme}

\begin{figure*}[bt]
    \centering
    \includegraphics[scale=0.6]{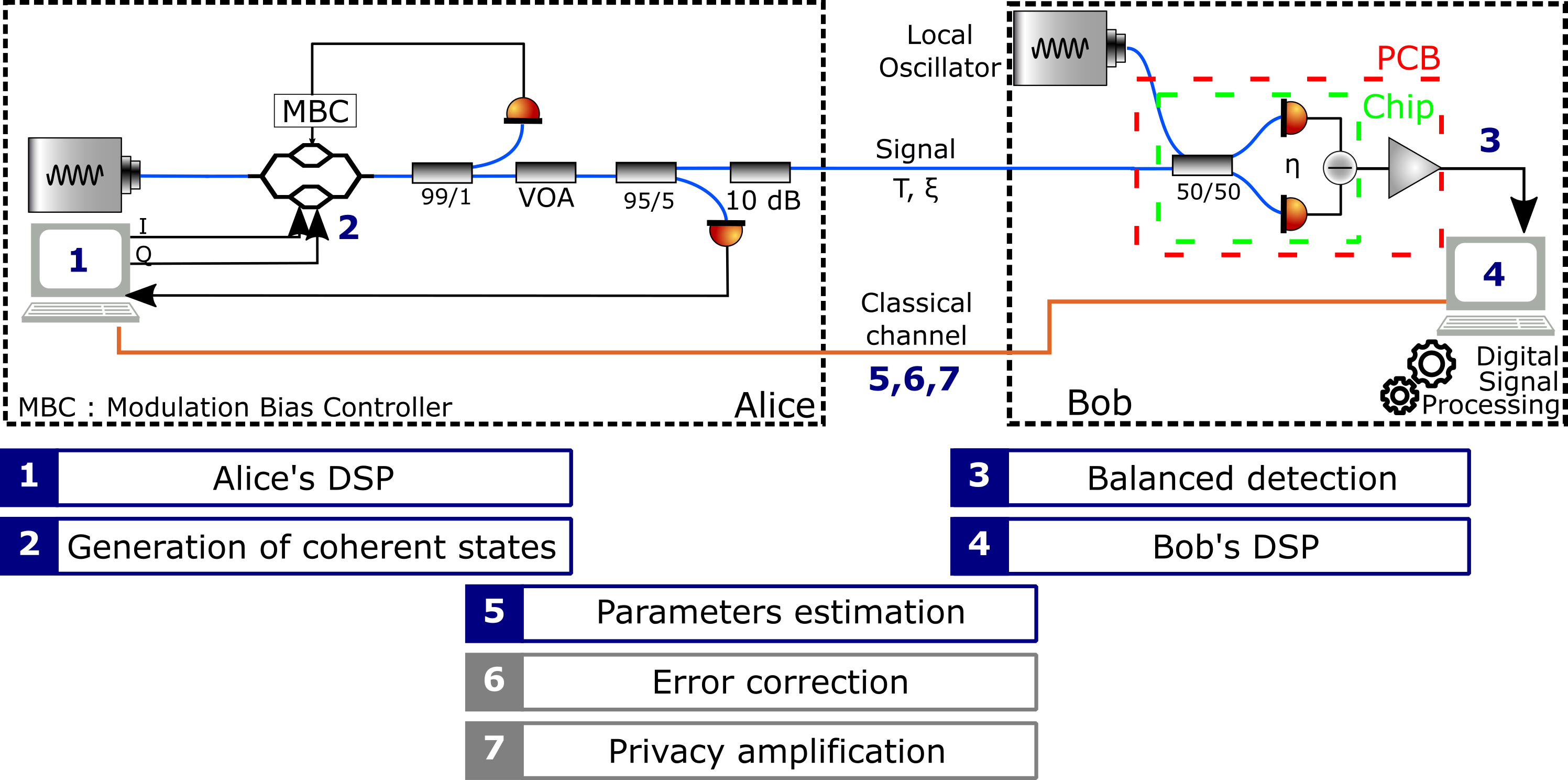}
    \caption{\textbf{Experimental scheme of our \gls{cvqkd} system.} In our experiment, the classical channel is a local ethernet link. The quantum channel is emulated by an electronic \gls{voa}. \gls{adc} and \gls{dac} are PCIe cards. An electronic VOA is used to tune the output power at Alice's site. In our experiment, steps 1 to 5 of the CV-QKD protocol are implemented.}
    \label{fig:experimental_scheme}
\end{figure*}

We implement a variant of the GG02 protocol~\cite{Grosshans_2002} using coherent states with \gls{ossb} modulation. The general steps of the protocol are summarized in Fig.~\ref{fig:experimental_scheme}. At step~\textbf{1}, Alice draws random symbols according to a predefined probability distribution (Gaussian, \gls{psk}, \gls{qam}, \gls{pcs}-\gls{qam}) and prepares the signal for the transmission by applying her \glsxtrfull{dsp}. At step~\textbf{2}, Alice generates the coherent states for the protocol by applying the signal to an IQ modulator and sends them to Bob through a quantum channel defined by its transmittance $T$ and excess noise $\xi$. At step~\textbf{3}, Bob detects the states using coherent detection, which has efficiency $\eta$ and electronic noise $\sigma_{el}^2$. At step~\textbf{4}, Bob applies his \gls{dsp} to the received signal to recover the symbols sent by Alice. More information on the \gls{dsp} can be found in subsection~\ref{subsec:dsp} and Fig.~\ref{fig:dsp}. At step~\textbf{5}, Alice and Bob use the classical channel and estimate the parameters needed to calculate the secret key rate, namely the modulation variance $V_A$, $T$ and $\xi$.
At step~\textbf{6}, they use the classical channel to perform reverse reconciliation, whereby Alice corrects her data to match Bob's data. Finally, at step~\textbf{7}, they perform privacy amplification. At the end of this step, Alice and Bob share a secret common bitstring. 

In this work, we implemented steps 1 to 5. While steps 6 and 7 are crucial for a real CV-QKD exchange since they allow the extraction of a secret key, their implementation is beyond the scope of this paper. More information on the related techniques can be found in~\cite{Gumus2021, Bai2022}.

The general schematic view of the optical setup is shown in Fig.~\ref{fig:experimental_scheme}. Alice is composed of a continuous wave laser (NKT Koheron Basik) that is used for the generation of the signal, which goes into an IQ modulator (Exail) controlled by its \gls{mbc}. The role of the \gls{mbc} is to act as a feedback loop to apply the required bias voltages to the modulator. After the \gls{mbc}, the path goes into an electronic VOA (Thorlabs VP1150A) and a 95/5 beam splitter, where 95\% of the light is measured on a photodiode to estimate Alice's modulation variance $V_A$, related to the mean number of photons emitted per symbol by $\langle n \rangle = \frac{V_A}{2}$. Finally, a fixed $10\,\si{\deci\bel}$ attenuator allows reaching the quantum levels required for \gls{cvqkd} ($\langle n \rangle \sim 1-3$ photons per symbol). Alice's \gls{dac} has $1\,\si{\giga\hertz}$ bandwidth and maximal sample rate $2\,\si{\giga\sample\per\second}$ (Teledyne SDR14Tx).

Fiber loss in the quantum channel is then emulated by an electronic and polarization maintaining \gls{voa} (Thorlabs V1550PA) that can attenuate the signal in a range from 0 to $31\,\si{\deci\bel}$, allowing for emulation of up to $155\,\si{\kilo\meter}$ of distance considering an attenuation of $0.2\,\si{\deci\bel\per\km}$ for the optical fiber.

On Bob's side, another continuous wave laser of the same type as the one at Alice's setup is used to generate the powerful \gls{lo}, in a so-called true local oscillator configuration. The \gls{lo} is mixed with the signal on chip in a 50/50 beam splitter and detected by the coherent detector. Bob's \gls{adc} works at up to $1.25\,\si{\giga\hertz}$ and at a maximal sample rate of $2.5\,\si{\giga\sample\per\second}$ (Teledyne ADQ32).

We recall here that in \gls{cvqkd} one usually works in \gls{snu}~\cite{Laudenbach_2018}, where the quantum noise has unit variance, which corresponds to choosing $\hbar=2$. This means that we normalize all measurements at Bob's side by the shot noise, and every noise variance will thus be given in \gls{snu}. At Alice's side, the normalisation is done through the average number of photons $\langle n \rangle$, as explained above, so that Alice and Bob evaluate their quantities in the same units.

Calibration is an important part of the \gls{cvqkd} protocol implementation since parameters typically vary. A first category of calibrations are made only once (or each time the optical setup is changed). This is the case for the calibration of the power ratio between the monitoring photodiode and the output of Alice (this will later allow us to estimate $\langle n \rangle$), the efficiency of the detector $\eta$ and the value of the electronic noise, which is assumed to be constant. A second category of calibrations are made between each \gls{cvqkd} frame. This concerns the estimation of the shot noise, which allows us to take possible power fluctuations of Bob's laser into account. The average number of photons $\langle n \rangle$ is also computed for each frame. Both the electronic noise and shot noise acquisitions are then processed using the same \gls{dsp} as for the quantum signal before being used to compute $\sigma_0^2$ and $\sigma_{el}^2$ (whose value can then slightly change at each round).

\subsection{Digital Signal Processing\label{subsec:dsp}}

\begin{figure}
    \centering
    \includegraphics[scale=0.38]{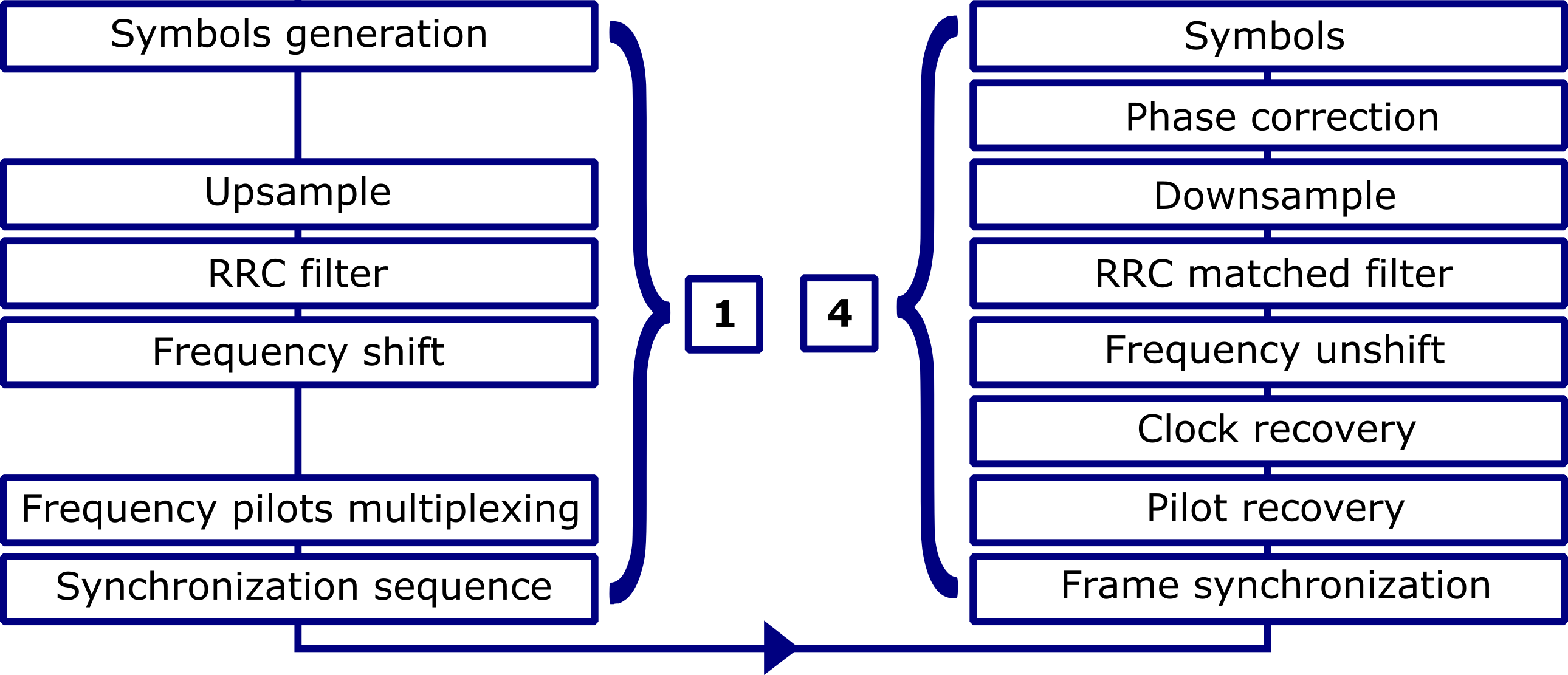}
    \caption{\textbf{Digital Signal Processing scheme.} Step 1 corresponds to the DSP on Alice's side, and should be read from top to bottom. Step 4 corresponds to the DSP on Bob's side and should be read from bottom to top.}
    \label{fig:dsp}
\end{figure}

\glsxtrlong{dsp} is the ensemble of the actions of generating the symbols and preparing them for physical transmission at Alice's side and receiving and recovering the symbols at Bob's side making use of digital filters and operations. We implemented a specifically designed \gls{dsp} for \gls{cvqkd} including the steps shown in Fig.~\ref{fig:dsp}.

On Alice's side, the symbols are generated using a modulation pattern and a source of entropy. Here a standard pseudo-random number generator was used as source of entropy but in a full implementation a \gls{qrng} would be used instead. The modulation can be chosen between Gaussian, \gls{psk}, \gls{qam} and \gls{pcs}-\gls{qam}. Note that although the use of discrete modulations presents significant practical advantages, with a potential of excellent performance~\cite{Roumestan_arXiv2022}, the security analysis of such protocols has only recently progressed with security proofs in the asymptotic case~\cite{Denys2021,Lin_2019}. After symbol generation, the sequence is upsampled and pulse shaped using a \gls{rrc} filter. This is done to minimize \gls{isi} and to optimize the transmission in the physical domain. The quantum symbols are then shifted in frequency, in order to reduce the excess noise coming from low frequencies and to be able to perform an RF-heterodyne, which allows to measure both quadratures with only one balanced detector; see appendix~\ref{appendix:optical_coherent_detection} for more details. Two pilot tones, which are complex exponentials, are then added to the signal at different frequencies. These tones are used to correct the clock difference, as a frequency reference and as a phase reference to correct phase noise coming from the transmission. Finally a Zadoff-Chu synchronization sequence is added at the beginning of the sequence.

\begin{figure}
    \centering
    \input{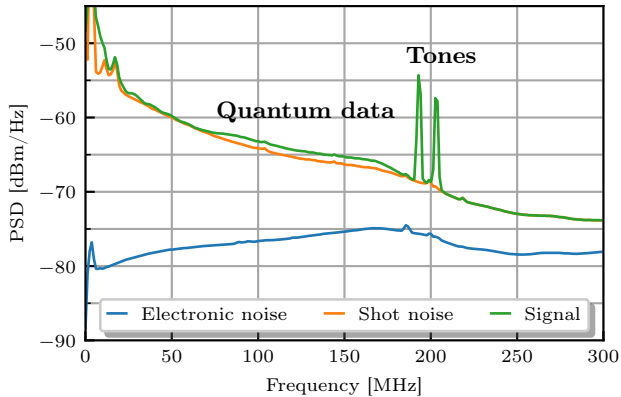}
    \caption{\textbf{Power Spectral Density of a representative received sequence.} The \gls{dsp} parameters are similar to those used in the CV-QKD experiment (see table \ref{tab:transmission_parameters}). The signal has a bandwidth of $130\,\si{\mega\hertz}$ and is centered at $125\,\si{\mega\hertz}$. The two pilots are at $190\,\si{\mega\hertz}$ and $200\,\si{\mega\hertz}$. The $2\,\si{\mega\hertz}$ frequency shift corresponds to the beat difference between the two lasers. Some noise can be seen at low frequencies and is filtered out during Bob's \gls{dsp}. For display purposes only, Alice's variance was set at around $62\,\si{\snu}$ in this plot. The frequency axis is shown here in linear scale for clarity.}
    \label{fig:recovered}
\end{figure}

On Bob's side, the signal is first detected; the Power Spectral Density of one received signal is shown as an example in Fig.~\ref{fig:recovered}. The \gls{dsp} at Bob's side goes as follows: the synchronization sequence is recovered by cross-correlating the received signal with the perfect Zadoff-Chu sequence. The frequencies of the two pilot tones are then recovered and the frequency difference allows to correct for a clock mismatch. Then the frequency of one of the pilots gives a frequency reference that is used for carrier frequency recovery. The quantum symbols are unshifted and filtered again using the same \gls{rrc} filter, which in total gives a \gls{rc} filter, that is known to minimize \gls{isi}. The sequence is then downsampled by finding the optimal sampling point where the variance of the received signal is maximal. The phase is corrected using the phase reference tones and the global phase is corrected by maximising the covariance between Alice's and Bob's data.

This \gls{dsp} scheme requires to set many parameters such as the symbol rate, the roll-off of the \gls{rrc} filter, the frequency of the reference pilot, and the frequency shift of the quantum symbols (among others). We specifically optimized the values of those parameters to reach the best performance %efficiency 
for the integrated receiver platform. The parameters are summarized in Table~\ref{tab:transmission_parameters}.

\begin{table}
    \centering
    \begin{tabular}{|c|c|}
        \hline
        \textbf{Parameter} & \textbf{Value} \\
        \hline
        Modulation & Gaussian \\
        \hline
        Rate ($R$)& $100\,\si{\mega\baud}$\\
        \hline
        Roll-off ($\beta$) & $0.3$\\
        \hline
        Bandwidth ($B$) & $130\,\si{\mega\hertz}$ \\
        \hline
        Freq. shift ($f_{\mathrm{shift}}$) & $125\,\si{\mega\hertz}$\\
        \hline
        Pilot tone 1 ($f_{\mathrm{pilot}, 1}$) & $190\,\si{\mega\hertz}$\\
        \hline
        Pilot tone 2 ($f_{\mathrm{pilot}, 2}$) & $200\,\si{\mega\hertz}$\\
        \hline
        Num. of symbols ($N$) & $10^6$ \\
        \hline
    \end{tabular}

    \caption{\textbf{Summary of the chosen DSP parameters for the \gls{pic}-based receiver.}  The power ratio between the pilot tones and the quantum signal is around $12\,\si{\deci\bel}$.}
    \label{tab:transmission_parameters}

\end{table}

The \gls{dsp} was tested to check that it was able to recover frames of symbols sent by Alice, before moving to parameter estimation.

\subsection{Excess noise estimation}

As we discussed previously, in \gls{cvqkd}, parameter estimation is an important step where three parameters are typically estimated: Alice's modulation strength $V_A$, the channel attenuation $T$ and the excess noise at Alice's output $\xi$. These parameters, along with the efficiency and the electronic noise of the receiver and the error correction efficiency, allow computing the secret key rate using the Devetak-Winter formula \cite{Devetak_2005}:
\begin{equation}
    k = \beta_{EC}I_{AB} - \chi_{BE},
\end{equation}
where $\beta_{EC}$ is the error correction efficiency, whose standard value in the literature is 95\%, $I_{AB} = \log_2(1+\text{SNR})$ is the maximal mutual information between Alice and Bob, and $\chi_{BE}$ is the Holevo bound on the information shared between Bob and Eve, where this formula is true in the reverse reconciliation setting where Bob's bit string is used to correct Alice's string to match it with his. The usual technique to estimate those parameters is to use covariance matrices~\cite{Laudenbach_2018}. The mean number of photons measured at Alice's monitoring photodiode allows finding the conversion factor and scaling Alice's sequence accordingly. This sequence is then sent to Bob, who computes the covariance matrix and finds:
\begin{equation}
\begin{split}
    T &= 2\cdot\frac{\langle XY \rangle^2}{\eta V_A^2}\\
    \xi &= 2\cdot\frac{\langle Y^2 \rangle - 1 - V_{el} - \frac{\eta T}{2} V_A}{\eta T},
\end{split}
\end{equation}
where $X$ and $Y$ respectively correspond to a subset of Alice and Bob symbols that are announced on the public channel and discarded.

\begin{figure}
    \centering
    \input{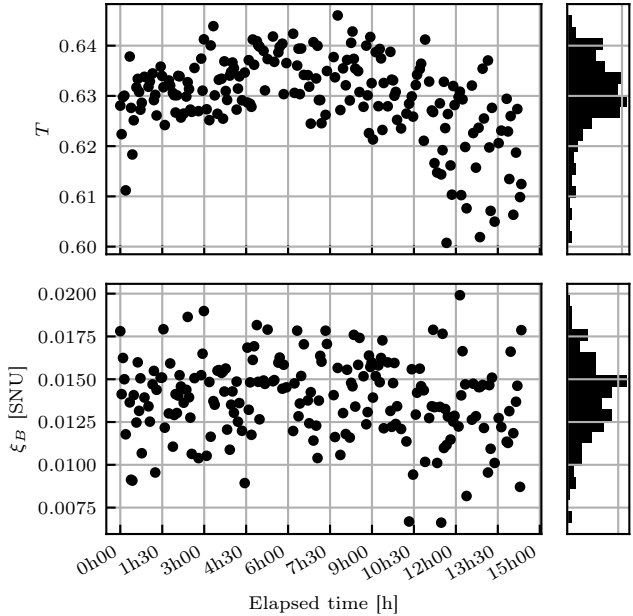}
    \caption{\textbf{Excess noise measurements for the successful frames over 15 hours on the receiver platform for the $10\,\si{\km}$ experiment.} The upper plot shows the evolution of $T$ assuming constant $\eta = 0.175$, while the lower plot shows the evolution of the excess noise $\xi_B$ (at Bob's side). On the right of each plot, the histogram of the data with 20 bins.}
    \label{fig:excess_noise_exp1}
\end{figure}

\begin{figure}
    \centering
    \input{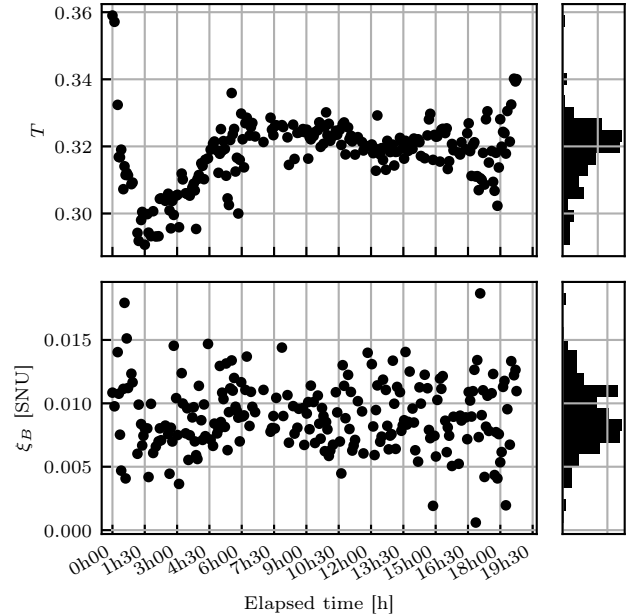}
    \caption{\textbf{Excess noise measurements for the successful frames over 18 hours on the receiver platform for the $23\,\si{\km}$ experiment.} The upper plot shows the evolution of $T$ assuming constant $\eta = 0.161$, while the lower plot is the evolution of the excess noise $\xi_B$ (at Bob's side). On the right of each plot, the histogram of the data with 20~bins.}    
    \label{fig:excess_noise_exp2}
\end{figure}

These steps were implemented after the \gls{dsp}. We performed two experiments for excess noise measurement, each experiment consisting of the acquisition of 300 CV-QKD frames and roughly $15\,\si{\hour}$ of acquisitions and \gls{dsp}. The first experiment was conducted with the \gls{voa} voltage set at $2.5\,\si{\volt}$ (theoretical attenuation of $2\,\si{\deci\bel}$ corresponding to $10\,\si{\kilo\meter}$ of equivalent distance) and the second one with a voltage of $2.9\,\si{\volt}$ (theoretical attenuation of $3.8\,\si{\deci\bel}$ corresponding to $19\,\si{\kilo\meter}$ of equivalent distance). The actual attenuation in the second experiment was measured to be slightly higher, at an equivalent distance of $23\,\si{\kilo\meter}$. The results of the two  experiments are shown in Figs.~\ref{fig:excess_noise_exp1} and \ref{fig:excess_noise_exp2}, respectively. Out of the 300 \gls{cvqkd} frames the \gls{dsp} failed 107 times for the first experiment and 82 times for the second experiment, respectively yielding 193 and 218 exploitable frames. These numbers are reflected in the \gls{fer} values given in Table~\ref{tab:parameters_estimation_cvqkd}.

The plots show the values of $T$ and $\xi_B = \eta T \xi$ at Bob's side during the measurements. Some fluctuations are clearly seen and we attribute them to physical vibrations slowly moving the fiber array away from the optimal coupling position. Indeed, although our coupling remains stable in time, it is sensitive to mechanical vibrations on the optical table and more generally in the experimental laboratory. A similar analysis (using the same software) with the same setup for Alice and a setup with bulk components for Bob yielded better stability performance. These fluctuations are definitely a challenge for this integrated receiver and indicate that fiber attachment would greatly benefit its operation.

Regarding the coupling, we also remark that the 26\% of efficiency obtained in the characterisation experiments was actually hard to reach in full system operation. The optical input that featured the best balance with the minimal attenuation on the \gls{voa}s was used for the local oscillator in the CV-QKD experiment to prevent saturation in the electronics and get the best efficiency on the signal input. This explains why the achieved efficiency was 17.5\% and 16.1\% respectively for the two experiments; see Table~\ref{tab:parameters_estimation_cvqkd} that summarizes the estimated experimental parameters.

Our experiments yielded an average value of $\xi_B$ of, respectively, $0.014\,\si{\snu}$ and $0.009\,\si{\snu}$ at Bob's side. As shown in Table~\ref{tab:parameters_estimation_cvqkd}, we have all the parameters necessary to estimate the \gls{cvqkd} rate.

\begin{table}[H]
    \centering
    \begin{tabular}{|c|c|c|}
        \hline
        \textbf{Parameter} & \textbf{10 km experiment} & \textbf{23 km experiment}\\
        \hline
         $V_A$ & $4.10\,\si{\snu}$ & $5.45\,\si{\snu}$ \\
        \hline
         $T$ & $0.632$ & $0.346$\\
        \hline
         $\xi_B$ & $0.014\,\si{\snu}$ & $0.009\,\si{\snu}$\\
        \hline
         $\eta$ & $0.175$ & $0.161$ \\
        \hline
         $V_{el}$ & $0.086\,\si{\snu}$ & $0.097\,\si{\snu}$\\
        \hline
         FER & $0.36$ & $0.27$\\
        \hline
        Asymptotic SKR & $2.4\,\si{\mega\bit\per\second}$ & $220\,\si{\kilo\bit\per\second}$\\
        \hline
    \end{tabular}
    \caption{\textbf{Average values of the estimated parameters for CV-QKD.} The average was calculated respectively on 193 frames and 218 frames for the $10$ and $23\,\si{\km}$ experiments.}
    \label{tab:parameters_estimation_cvqkd}
\end{table}

For the first experiment, taking the average values of the excess noise and of the transmittance yields a secret key rate (SKR) of $2.4\,\si{\mega\bit\per\second}$ in the asymptotic case. It is also possible to compute the key rate for each frame, with each of the 193 frames giving a positive key rate, also averaging to $2.4\,\si{\mega\bit\per\second}$.

These calculations were performed in the asymptotic regime, which means that we make the assumption that we can perfectly estimate $\xi$ and $T$ and that we do not have leakage in the privacy amplification step (we still however consider the error correction efficiency). There are several methods to account for finite-size effects and one of them is to compute the variance of $\xi$ and $T$ over all the frames and then the worst case estimators $T_{min}$ and $\xi_{max}$ such that probability of $\xi_{real} > \xi_{max}$ (respectively $T_{real} < T_{min}$) is less than some $\varepsilon$ (usually $10^{-10}$). However this method here does not work mainly because the variations in the estimated transmittance and excess noise are high due to the unstable coupling. Instead, we conducted an analysis considering finite-size regime using the method in~\cite{Leverrier_2010}, with the same security parameters. We also take into account the fact that only half of the symbols are actually used for the key generation process while the other half is used for parameter estimation. The \glsxtrfull{fer} is not taken into account in these calculations.

\begin{figure}
    \centering
    \input{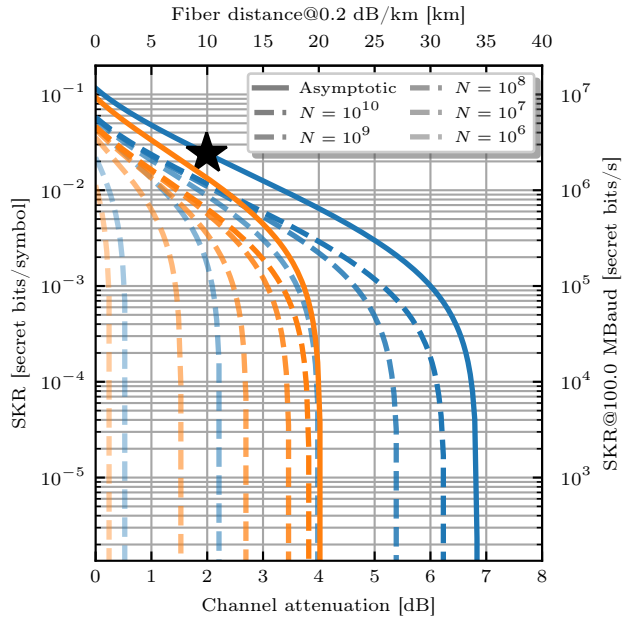}
    \caption{\textbf{Secret key rate estimation.} The blue and orange lines correspond to the $10\,\si{\kilo\meter}$ and $23\,\si{\kilo\meter}$ experiments respectively.}
    \label{fig:skr}
\end{figure}

The results for the first experiment are plotted in blue in Fig.~\ref{fig:skr}. The plain lines correspond to the asymptotic case and the dashed lines to different level of finite-size scenarios, \textit{i.e.}, the number of symbols that are considered for the parameter estimation.

We notice that the number of symbols used in our experiment ($10^6$) is not sufficient to obtain a positive key rate. Expected SKR values for $N=10^7, 10^8, 10^9$ and $10^{10}$ symbols are respectively $0.17, 0.88, 1.10$ and $1.18\,\si{\mega\bit\per\second}$ but such a number of symbols is challenging for the \gls{dsp} and the stability of the system, in particular since, as the \gls{dsp} will require more time to be executed, the frames would be more separated temporarily.

For the second experiment, computing the SKR from the average values gives a null result in the asymptotic case as can be seen from the plain orange line in Fig.~\ref{fig:skr}; in particular, we see that it falls at zero at $20\;\si{\kilo\meter}$ using the average values of $\xi$ and $T$. However, we can make the analysis by averaging the frames where the key rate was positive, asymptotic or with finite-size effects. Out of the 218 frames, 99 yielded a positive key rate and taking the average of these values yields a rate of $485\,\si{\kilo\bit\per\second}$, while averaging over all 218 frames gives $220\,\si{\kilo\bit\per\second}$.

We can also compute the finite-size key rate and the frames that gave a positive excess noise and we get the following results: for $N=10^7, 10^8, 10^9$ and $10^{10}$ we had among the 218 frames, respectively $1, 28, 70$ and $94$ frames that gave positive SKR values yielding an average (over the frames with a positive key rate) of $232, 199, 217$ and $221\,\si{\kilo\bit\per\second}$, while averaging over all the frames, we obtain $1, 26, 70$ and $96\,\si{\kilo\bit\per\second}$.

We also remark that with the parameters used in the $10\,\si{\kilo\meter}$ experiment, we would expect to be able to reach a range greater than $20\,\si{\kilo\meter}$. However this would only be true if all the parameters were constant (which is not the case for $\eta$ for instance) and if the excess noise was indeed coming entirely from the channel (which again is not the case, for instance the \gls{dsp} is less efficient at lower SNR). We remark as well that the excess noise increases when Alice's variance increases which makes it harder to find the optimal value for the number of photons at Alice's side. In our experiment we made, for each attenuation, a sweep for Alice's variance and took the value that was giving the best key rate on average.

\section{Conclusion\label{sec:concl}}

In this work, we presented a coherent receiver platform based on a Si photonic integrated device with promising intrinsic characteristics for \gls{cvqkd}. The measured excess noise was in a valid range for \gls{cvqkd} exchange and an asymptotic secret key rate in the order of hundred of $\si{\kilo\bit\per\second}$ was computed for attenuation corresponding to metropolitan distances. The receiver is based on a technology that is CMOS compatible, simple and can achieve scaling with low size and low cost.

The performance of the experiment is currently limited by a few factors including the packaging of the chip, especially the optical packaging that limits the efficiency of the receiver, by the electronic chain that can be improved to obtain a higher bandwidth and hence symbol rate, and by the \gls{dsp} process, where both the memory consumption and computation time can be optimised. The limitations are however not intrinsic to the photonic circuit, thus our results constitute a strong proof-of-principle for \gls{pic}-based \gls{cvqkd}.

Future versions of integrated receiver platforms could benefit from additional components, including a switch (or fast \gls{voa}) on the signal path, to block light during shot noise calibration, on-chip laser generation by using other PIC platforms such as InP~\cite{Aldama_2022}, direct $90^\circ$ mixing for phase-diverse heterodyne detection, and better optical packaging, which is critical for stability and efficiency. On the system level, future perspectives include full \gls{cvqkd} transmission, with implementation of error correction and privacy amplification, and experiments with a PIC-based transmitter~\cite{Aldama_2023} and receiver.

\section{Acknowledgments}
The authors acknowledge fruitful discussions with the partners of the concluded European Quantum Technologies Flagship project CiViQ. ED, LV and PG deeply thank Delphine Marris-Morini, Mauro Persechino and Melissa Ziebell for essential contributions at early stages of this work~\cite{Ziebell:15} and Jean Marc Fédéli from CEA-Leti for circuit fabrication. We acknowledge financial support from the European Union’s Horizon Europe research and innovation program under the grant agreement No 101114043 (QSNP).

\bibliography{bibliography.bib}
\appendix
\section*{Supplementary material}

\section{Amplification chain circuit design\label{appendix:TIA}}

The architectures that we considered for the TIA of our PIC were based on RF amplifiers, mainly from Texas Instruments (TI). The design methodology involved AC/DC small-signal and noise analysis using SPICE models, Monte-Carlo simulations, then lab validation on-PCB to confirm the performances using scopes and spectrum analyzers. The electronics was tested alone using electronic inputs and with a bulk \gls{bhd} made of photodiodes from Hamamatsu (G9801-22) before a final test with the \gls{pic}. We noticed for instance, that the RF amplifier OPA855, which offers a high gain bandwidth product ($8\,\si{\giga\hertz}$) and a low voltage input voltage noise ($0.98\,\si{\nano\volt/\sqrt{\hertz}}$) has an important contribution to the total output noise of the detector. This is due to its bipolar input transistors suffering from a high input current noise, which increases with frequency. Also, managing the stability of the OPA855 in a closed-loop topology was not trivial and required a coarse tuning of its (already) large feedback capacitor, thus limiting the frequency bandwidth. We also considered the FET input OPA858 as an alternative solution to the OPA855 due to its interesting announced noise performance, but it is intended for operations in single-supply (e.g. time flight measurement with a single photodiode). Our investigation converged finally to the OPA818 to be a good candidate. It has FET input transistors and could operate at both single ($10\,\si{\volt}$) and dual ($-5\,\si{\volt}$/$+5\,\si{\volt}$) supplies. Despite its lower gain-band product ($2.7\,\si{\giga\hertz}$) compared to the OPA855 and to the OPA858, it achieves a very low input current noise $3\,\si{\femto\ampere/\sqrt(Hz)}$ at $10\,\si{\kilo\hertz}$ ($3\,\si{\pico\ampere/\sqrt{\hertz}}$ at $100\,\si{\mega\hertz}$) and offers a low common-mode input capacitance ($1.9\,\si{\pico\farad}$) and a wide dynamic output range allowing flexibility for amplification gain setting.

To optimize the amplification chain circuit shown in Fig.~\ref{fig:amplification_circuit}, 
we set the \gls{tia} gain to $R_f = 10\,\si{\kilo\ohm}$ to reach a closed-loop $Fc_{-3dB}$-bandwidth of about $150\,\si{\mega\hertz}$ in simulations for a maximum detector capacitance of $5\,\si{\pico\farad}$. The non-inverting gain was set to $R_2/R_1 +1 = 11$ offering a total chain gain (I to V) of $110\;\si{\kilo\volt\per\ampere}$ over the voltage output range of $V_+ = 5\,\si{\volt}$, $V_- = -5\,\si{\volt}$. $C_f$ was first determined by simulation regarding the estimated maximum detector capacitance (parasitic, \gls{tia}'s input and junction capacitors). The total parasitic capacitor is minimized due to appropriate PCB routing skills, by improving electrical bonding and selecting small packages (QFN, WSON,...). The photodiode's junction capacitor depends on the manufacturing process, and its value is inversely proportional to the applied reverse voltage. In our case, under $500\,\si{\milli\volt}$, the equivalent Silicon-Germanium \gls{bhd} capacitor (two parallel junction capacitances) is estimated to be of some dozens of femto-Farad and by far, not the limiting factor of the \gls{tia}'s frequency bandwidth. Indeed, the feedback capacitance estimated by simulation, requires a review and a fine tuning on PCB to achieve the largest bandwidth while ensuring system stability versus the detector total equivalent capacitance. 

\section{Optical coherent detection\label{appendix:optical_coherent_detection}}

A Balanced Homodyne Detector (BHD) can be seen as an amplified quadrature measurement, and usually the term  \textit{homodyne optical detection} or simply \textit{homodyne} is used, at least in \gls{cvqkd}, to describe the detection of one quadrature, which can be done with one balanced detector.

On the other hand, the term \textit{heterodyne} refers, in \gls{cvqkd}, to the simulaneous measurement of the two quadratures. The more straightforward way to do this is by using a 90$^\circ$ instead of a 180$^\circ$ mixer. This is sometimes referred to as \textit{phase-diverse heterodyne}. This however requires two \gls{bhd}s and extra steps in the \gls{dsp} (such as equalization for instance).

The technique of measuring both quadratures with only a single balanced detector is called \textit{RF-heterodyne}. It consists of making a frequency displacement of the signal by a frequency $f_c$. To ensure however that there is no overlap in the frequency spectrum, one has to impose that $f_c > \frac{B}{2}$, where $B$ is the bandwidth of the signal. Also, one has to ensure that only one sideband is modulated on Alice's side; improper sideband suppression can lead to errors and information leakage creating a side channel~\cite{Hajomer_2022}. This also means that half of the bandwidth is not available and the use of phase-diverse heterodyne would allow a doubling in the \glsxtrshort{skr}.

The RF-heterodyne method was used in the experiments reported in the present work, but we are also investigating the use of multiple \gls{bhd}s, either to perform phase diversity heterodyne and/or polarisation diversity detection, notably with a power supply board capable of powering up to 4 cards and with the additional required external discrete components. This would however also require a custom fiber array or several coupling stages.
\end{document}